\documentclass[conference,onecolumn,draftclsnofoot]{IEEEtran}

\IEEEoverridecommandlockouts
\usepackage{cite}
\usepackage{amsmath,amssymb,amsfonts}
\usepackage{algorithmic}
\usepackage{graphicx}
\usepackage{textcomp}
\usepackage{xcolor}
\def\BibTeX{{\rm B\kern-.05em{\sc i\kern-.025em b}\kern-.08em
    T\kern-.1667em\lower.7ex\hbox{E}\kern-.125emX}}
    
\usepackage[justification=justified,font=footnotesize,skip=0pt]{caption}
\allowdisplaybreaks

\usepackage{optidef}

\begin{document}

\title{Electrical Vehicle Fleet Routing Accounting for Dynamic Battery Degradation
}


\author{\IEEEauthorblockN{Daniel Gebbran}
\IEEEauthorblockA{\textit{Department of Electrical Engineering} \\
\textit{Technical University of Denmark}\\
Kgs. Lyngby, Denmark \\
dgeco@elektro.dtu.dk}
\and
\IEEEauthorblockN{Jeppe Rich}
\IEEEauthorblockA{\vspace{-2mm}\textit{Department of Technology,}\\
\textit{Management and Economics} \\
\textit{Technical University of Denmark}\\
Kgs. Lyngby, Denmark \\
rich@dtu.dk}
\and
\IEEEauthorblockN{Tomislav Dragi\v{c}evi\'{c}}
\IEEEauthorblockA{\textit{Department of Electrical Engineering} \\
\textit{Technical University of Denmark}\\
Kgs. Lyngby, Denmark \\
tomdr@elektro.dtu.dk}
}



\maketitle

\begin{abstract}
The increasing uptake of electrical vehicles (EVs) has increased the awareness of battery degradation costs and how they can be minimized. However, from a planning perspective it is difficult to integrate battery degradation models into existing route planning models and to assess how policies that aim at reducing battery degradation affect route planning costs and degradation across the fleet. In this paper, a simple transportation vehicle routing problem (VRP) is formulated as a mixed-integer nonlinear problem (MINLP), with a modification that allows monitoring the maximum and minimum depth-of-discharge (DoD) of the entire fleet. 
This allows us to measure the battery health degradation \emph{during} the online optimization process. 
The results show that accounting for the impact of different route characteristics on battery degradation can have an impact on the route planning of the entire fleet as well as the battery degradation for all vehicles. The latter is achieved by forcing vehicles to adapt to certain DoD boundaries in the long term.
\end{abstract}

%
%
%


\begin{IEEEkeywords}
Vehicle route problem, EV fleet, battery degradation, depth-of-discharge, cycle lifetime.
\end{IEEEkeywords}

\section{Introduction} \label{ch:Intro}
The electrification of transport will be an essential part of addressing goals for reduction in $\textnormal{CO}_2$ emissions since transportation accounts for 32\% of final energy demand, of which 96.7\% is supplied by non-renewable energy as of 2017 \cite{REN21_2020}. 
While electrical vehicles (EVs) accounted for only 2.6\% of global car sales in 2019, they registered a 40\% year-on-year increase, and electrification of transport is gaining momentum by ambitious policy initiatives, which define clear, long-term signals for the auto industry \cite{IEA_Global_EV_2020}. 
In the past few years, government entities and car manufacturers have all made commitments and investments with respect to EV technology. Furthermore, large corporations are in the process to transfer their vehicle fleets to EV fleets and non-automotive companies are investing more heavily in the sector \cite{REN21_2020}. 

This increasing uptake of EVs is also altering the way that vehicle routing strategies are computed, which now has to account for both the traditional routing problem combined with characteristics arising from EV operation. Therefore, new factors come into play, such as the charging strategies \cite{Yu_James_2018} and infrastructure\footnote{While optimal placing and sizing of charging infrastructure for EVs is another important topic, it is in itself, it constitute yet another research problem. For more details, please refer to \cite{Jensen_Rich_2021} and references therein.}, 
battery range, battery degradation, speed- and acceleration-dependent power consumption \cite{Dias_Cachinero_2020}, and the altered waiting times at recharging stations caused by the longer charging time when compared to fueling internal combustion vehicles (ICVs) \cite{Jensen_Rich_2021}. 

In parallel, recent research has also focused on how to address such factors, analyzing driving techniques for predictive energy management of EVs \cite{Zhou_Yang_2019}, or dynamic recommendations based on driving patterns \cite{Eider_Markus_2018}. However, even the most recent works 
\cite{Dias_Cachinero_2020} only addresses the degradation modeling within routing problems to a limited extent.

In contrast, the actual use of more accurate battery models -- including degradation -- in optimization frameworks can bring complexities in which they invoke the use of non-deterministic (non-differentiable) functions \cite{Lee_Jin_2022}. 
The battery degradation can be modeled based on thermodynamics and battery state-of-charge (SoC) dynamics, both of which affect cycle degradation and calendar aging, finally reflecting the difference in battery performance and total loss of capacity \cite{Calearo_Degrag_2019}. The mission profile (i.e., static and dynamic parameters and variables) of EVs operation, determining temperature and power profile can, therefore, be used to compute the battery degradation. In particular, different profiles for charging and discharging the battery can determine variations in the battery depth-of-discharge (DoD), which in turn affects advanced cycles. A good method to account for the degradation is the rainflow-counting algorithm, which also reflects the fact that higher DoDs can lead to a more pronounced battery degradation \cite{Schneider_Simon_2021}.


As such, recent works consider degradation models using approximations. In \cite{Dias_Cachinero_2020} it is proposed to use an operational planning model for route and charging of an EV fleet, for which the cost of battery degradation is included in the objective function. Its degradation model, however, is simplified in the sense that they use a general degradation cost based on average energy consumption (a parameter set \emph{a priori}) across vehicles, regardless of their DoD. 
In contrast, the use of more accurate degradation models within optimization frameworks will require the use of non-linear and possibly non-differentiable frameworks. Recent works have addressed partially this issue using a novel cost formulation proposed in \cite{Lee_Jin_2022}, where authors simplify the rainflow-counting algorithm to account for the discharging part of the battery operation, using an auxiliary one-cycle cost curve and a mathematical reformulation of the DoD with a single time delay to avoid non-differentiable logical expressions that cannot be interpreted by optimization solvers. 

Finally, if the transport sector is to get on track to meet global climate targets for 2030 goals and beyond, the momentum of transportation electrification has to be progressively increased \cite{IEA_Global_EV_2020}. Therefore, it is important to address open questions relating to the use of EVs and route planning algorithms. Given the previously described literature gap, the main contribution of this paper is the proposal of a framework that accounts for battery degradation \emph{dynamically} (i.e., preemptively) when calculating optimal routing strategies for a fleet of EVs. While it still uses a linear approximation of the DoD such as \cite{Dias_Cachinero_2020}, it assigns a cost associated with the difference among DoD across multiple vehicles in the problem, avoiding that any one vehicle has a higher DoD that could result in increased degradation over time, leading to a result similar to \cite{Lee_Jin_2022}. 



%
\section{Modeling approach} \label{ch:Model}
%

\subsection{Modified vehicle routing problem}

For any number of EVs $k \in \mathcal{K}$ and path between nodes $i=1,...,n$ and $j=1,...,n$, we formulate the following modified static VRP as a mixed-integer nonlinear problem (MINLP): 

\begin{subequations} \label{eq:VRP_Total}
\small
\begin{align}
&\hspace{-0.65cm} \underset{\boldsymbol{x} \in \{ 0, 1\}} {\mbox{minimize}} \quad  \sum_{k=1}^{ \mathcal{K}} \sum_{i=1}^{n} \sum_{j=1}^{n} \left( x_{ijk} c_{ij} + \alpha \: f(\boldsymbol{DoD}^{\textrm{fleet}}) \right), \label{eq:Cost_VRP}\\
\text{subject to:} \hspace{0.55cm} & \sum_{i=1}^{n} x_{ijk} = \sum_{i=1}^{n} x_{jik} \quad \forall \; j \in \mathcal{N}, k \in \mathcal{K} \label{eq:each_v_leaves_node_entered},\\
& \hspace{0.75cm} \sum_{k=1}^{p} \sum_{i=1}^{n} x_{ijk} = 1 \quad \forall \;  j \in \mathcal{N} \backslash \{1\} \label{eq:every_node_enter_once},\\
& \hspace{1.25cm} \sum_{j=2}^{n} x_{1jk} = 1 \quad \forall \;  k \in \mathcal{K} \label{eq:all_v_leave_depot},\\
& \hspace{-1.75cm} u_i - u_j + x_{ijk} * n \leq n - 1 \quad \forall \; i \in \mathcal{N} \backslash \{1\}, j \in \mathcal{N}, i \neq j, k \in \mathcal{K} \label{eq:MTZ},\\
& \hspace{0.25cm} \overline{DoD}^{\textrm{fleet}} \leq 100 - SoC^{\textrm{end}}_{k} \quad \forall \; k \in \mathcal{K \label{eq:DoD_max}}, \\
& \hspace{0.25cm} \underline{DoD}^{\textrm{fleet}} \geq 100 - SoC^{\textrm{end}}_{k} \quad \forall \; k \in \mathcal{K \label{eq:DoD_min}}, \\
& \hspace{-0.65cm} SoC^{\textrm{end}}_{k} = SoC^{\textrm{start}}_{k} - \sum_{i=1}^{n} \sum_{j=1}^{n} x_{ijk} * e_{ij} \quad \forall \; k \in \mathcal{K} \label{eq:SoC},
\normalsize
\end{align}
\end{subequations}
where $x$ is the binary variable representing whether vehicle $k$ is on the path between $i$ and $j$. $c$ is the travel cost associated with such path and \eqref{eq:Cost_VRP} is the total cost function which include a penalization term that account for battery degradation. \eqref{eq:each_v_leaves_node_entered} ensures each vehicle leaves the node it entered, \eqref{eq:every_node_enter_once} ensures every node is entered at least once, \eqref{eq:all_v_leave_depot} ensures all vehicles leave the depot $i=\{1\}$, \eqref{eq:MTZ} is the Miller-Tucker-Zemlin (MTZ) sub-tour elimination constraint, which ensures there are no sub-tours formed within each vehicle's tour, and uses the extra auxiliary variable $u$ \cite{Toth_2014_VRP}, $\alpha$ is the multi-objective weighting factor related to the associated degradation cost, $\overline{DoD}^{\textrm{fleet}}$ and $\overline{DoD}^{\textrm{fleet}}$ are the maximum and minimum DoD of the vehicles in the fleet, which is ensured by \eqref{eq:DoD_max} and \eqref{eq:DoD_min}, and $SoC^{\textrm{end}}_{k}$ is the final battery state-of-charge for each vehicle, which is computed using \eqref{eq:SoC}, using the energy $e_{ij}$ associated with each path and the vehicle's initial SoC, given by $SoC^{\textrm{start}}_{k}$.

The relaxation of the objective function, by adding the battery degradation penalization to the original VRP can be modeled in two different ways. The most intuitive way is to reduce the absolute difference between maximum and minimum DoDs for the EV fleet, using a quadratic term 
$f(\boldsymbol{DoD}^{\textrm{fleet}}) =  (\underline{DoD}^{\textrm{fleet}} - \overline{DoD}^{\textrm{fleet}})^2$. This will be referred to as \textit{Algorithm 1} and requires that the problem is stated as a mixed-integer nonlinear problem (MINLP). 
A linear simplification with $f(\boldsymbol{DoD}^{\textrm{fleet}}) = - \overline{DoD}^{\textrm{fleet}}$ allows the problem to be stated as a mixed-integer linear problem (MILP), referred to as \textit{Algorithm 2}.


%
\subsection{Battery degradation modeling}

\begin{figure}[htpb]
\begin{center}
\includegraphics[width=0.6\textwidth]{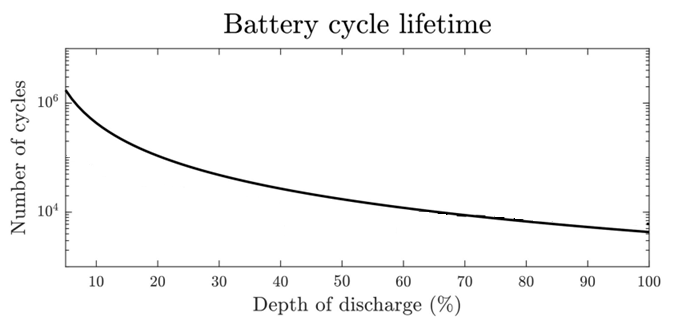}
\caption{Wôhler curve associating battery degradation with number of cycles and operational DoD \cite{Schneider_Simon_2021}.}
\label{fig:BatteryLifetime}
\end{center}
\end{figure}

In the paper, we apply a degradation model that accounts for the relationship between the DoD and the battery cycle lifetime. An example of a Wôhler curve is shown in Fig. \ref{fig:BatteryLifetime}, where the relationship between higher DoDs and battery degradation is shown by the reduction in the battery's lifetime.

\section{Results} \label{ch:Results}
The given VRP \eqref{eq:VRP_Total} has been modelled in Python, using Pyomo \cite{Pyomo2017Hart} and solved with MindtPy \cite{MindtPy}, which also makes use of IPOPT \cite{IPOPT}. The test case has used three vehicles $k \in \mathcal{K}$ and eight nodes for $i$ and $j$. Table \ref{table:results} depicts the results obtained using different weights%
\footnote{Changes in $\alpha$ between the given values shown in the table did not generate any different results.} %
for $\alpha$ for both \textit{Algorithm 1} and \textit{Algorithm 2}. The table illustrates the cost deviation in comparison to the basic VRP, the individual final SoC percentage, the DoD percentage, and the lifetime cycle estimation of the worst-case battery in the fleet, based on Fig. \ref{fig:BatteryLifetime}.
%
%

It is possible to see that without any change in the route cost, routes and charging strategies that lead to increasing discharging percentages, will cause severe degradation of the battery. However, assigning a large degradation weight might cause largely similar DoD levels across the fleet, which will also reduce the flexibility and the cost from the perspective of the operator. Still, this is a known property in multiobjective optimization problems \cite{Gebbran_APEN}.






This effect is especially notable in the management of EV fleets within large networks and extensive routes because the degradation is accrued over the DoD, which is affected in a short-term planning horizon. For instance, a given vehicle using routes with DoDs of 65\% and 20\%, every other day, will lead to a cycle lifetime under 10000 cycles (i.e., as seen in the result with $\alpha = 0$), whereas the same vehicle using routes averaging below 45\% DoD will (i.e., as seen in the result with $\alpha = 0.05$) have a lifetime over 25000 cycles -- effectively, more than doubling its cycle lifetime. On the contrary. traditional ICVs maintenance and mechanical degradation are accumulated over longer planning horizons, instead of differences in daily use.


\begin{table}[bp]
\caption{Simulation results}
\renewcommand{\arraystretch}{0.9}
\begin{center} \label{table:results}
\begin{tabular}{lccccc}
                & Base case    & \multicolumn{2}{c}{Algorithm 1} & \multicolumn{2}{c}{Algorithm 2}  \\
$\alpha$           & 0            & 0.05         & 0.125            & 0.05         & 2                 \\
Cost diff. [\%] & N/A          & 0            & 45               & 0            & 45                \\
$SoC_k^{\textrm{end}}$~[\%]     & \{34,79,80\} & \{59,75,59\} & \{41,43,44\}     & \{59,75,59\} & \{41,43,44\}      \\
$DoD_k$~[\%]     & \{66,21,20\} & \{41,25,41\} & \{41,43,44\}     & \{41,25,41\} & \{41,43,44\}      \\
Cycles          &  \textless{}10000       & \textless{}30000        & \textless{}27500 & \textless{}30000        & \textless{}27500            
\end{tabular}
\end{center}
\end{table}

\vspace{-2mm}
\section{Discussion} \label{ch:Discussion}
While there have been recent works that focus on the degradation of batteries, as discussed throughout the introduction, there is still much room for associating the dynamic degradation of EV batteries within VRPs. Using the present work as a starting point, this section discuss ramifications and directions for future research.


Intuitive extensions of the work herein presented include increasing the dimension of the problem, temporally and spatially, including more diverse simulation scenarios. Moreover, accounting for the charging schedules of the fleet of EVs \cite{Dias_Cachinero_2020} and associating the SoC of EVs across time periods are natural extensions. On top of that, combining predicted vehicle demand forecast and prioritizing the charging of EVs during periods with lower electricity prices are other improvements that have been investigated in the literature, but without considering the dimension of degradation effects in the fleet as herein presented. Naturally, those modifications require adopting more advanced algorithms instead of the simple MINLP herein adopted.

Furthermore, the problem can be re-cast within the purview of on-demand services, which allows for a more truthful representation of reality where a differentiated organization for charging and operation intervals can be associated with predicted fleet autonomy, charging time, and expected service requests over a given time horizon.

With regards to battery technologies, an improved rainflow-counting algorithm with included cell temperature effects and specifics for specific fleet battery cells 
can be implemented similarly as \cite{Lee_Jin_2022}, which allows for a more detailed calculation of degradation during the operation of the fleet (e.g., by improving the association between the degradation cost $\alpha$). In parallel, more detailed real-time data acquisition can lead to a finer degree of degradation modeling, associating thermodynamic impacts on the battery \cite{Calearo_Degrag_2019} with aggressive driving patterns, and even extending the impact of such driving behaviors associated with loss of energy (e.g., reduced use of regenerative braking, and frequent intense acceleration episodes) -- which has been found to account for nearly 40\% reduction in battery life when compared to gentle drivers \cite{Jafari_Mehdi_2018}. These behaviors can also be identified using outlier detection with online or post-processing diagnosis. 
Moreover, the degradation of the battery itself should not be taken as a sole decision variable and must be combined, for example, with the total energy expenditure and the total transportation cost having in mind the actual degradation cost in comparison to these other quantities.

Finally, because calendar aging also plays an important role, it is necessary to re-evaluate the fleet dispatch accordingly. In other words, excessive degrading of a single EV compared to others will inevitably lead to a battery replacement, associated with high costs, whereas the cycles "saved" by the other batteries will in fact never be used, since the battery's calendar aging itself will then be the pronounced cause for the loss of capacity \cite{Calearo_Degrag_2019}. Moreover, an investigation on the use of second-hand use batteries might prove a profitable investment for fleet operators. For instance, if the known-ahead operating conditions have unavoidable high cycle degradation, the reduced cycle lifetime will be compensated by the lower purchase costs of second-hand batteries, and still be a good investment even with a smaller calendar lifetime \cite{Schneider_Simon_2021}.






\vspace{-4mm}
\section{Conclusion} \label{ch:Conclusion}
The electrification of transport prompts for improved algorithms that account for the characteristics of electrical vehicles (EVs) to best use their potential and increase their operational efficiency. A modified version of the transportation vehicle routing problem (VRP) was formulated, accounting for the dynamic depth-of-discharge (DoD) on different EVs of an operating fleet. It has been shown that by accounting for battery degradation across the fleet, by forcing the difference between DoD levels for the fleet to be small, battery degradation can be reduced significantly and thus prolonging the cycle lifetime of the batteries, while not negatively affecting the fulfillment of logistic demands. This has been shown in a simple example network, and the full effect of the battery degradation was computed in retrospect. 
Unlike other recent works, this allows a simple way of accounting for the battery degradation across the fleet of EVs. Finally, other relevant issues have been discussed, describing possible future directions for this research area.





\bibliographystyle{IEEEtran}
\bibliography{MainText}{}

\begin{thebibliography}{10}
\providecommand{\url}[1]{#1}
\csname url@samestyle\endcsname
\providecommand{\newblock}{\relax}
\providecommand{\bibinfo}[2]{#2}
\providecommand{\BIBentrySTDinterwordspacing}{\spaceskip=0pt\relax}
\providecommand{\BIBentryALTinterwordstretchfactor}{4}
\providecommand{\BIBentryALTinterwordspacing}{\spaceskip=\fontdimen2\font plus
\BIBentryALTinterwordstretchfactor\fontdimen3\font minus
  \fontdimen4\font\relax}
\providecommand{\BIBforeignlanguage}[2]{{%
\expandafter\ifx\csname l@#1\endcsname\relax
\typeout{** WARNING: IEEEtran.bst: No hyphenation pattern has been}%
\typeout{** loaded for the language `#1'. Using the pattern for}%
\typeout{** the default language instead.}%
\else
\language=\csname l@#1\endcsname
\fi
#2}}
\providecommand{\BIBdecl}{\relax}
\BIBdecl

\bibitem{REN21_2020}
{REN 21}, ``{Renewables 2020 Global Status Report (GSR)},'' Tech. Rep., 2020.

\bibitem{IEA_Global_EV_2020}
{IEA}, ``{Global EV Outlook},'' Tech. Rep., June 2020.

\bibitem{Yu_James_2018}
J.~J.~Q. Yu and A.~Y.~S. Lam, ``Autonomous vehicle logistic system: Joint
  routing and charging strategy,'' \emph{IEEE Transactions on Intelligent
  Transportation Systems}, vol.~19, no.~7, pp. 2175--2187, 2018.

\bibitem{Jensen_Rich_2021}
A.~F. Jensen, M.~Thorhauge, S.~E. Mabit, and J.~Rich, ``Demand for plug-in
  electric vehicles across segments in the future vehicle market,''
  \emph{Transportation Research Part D: Transport and Environment}, vol.~98, p.
  102976, 2021.

\bibitem{Dias_Cachinero_2020}
P.~Diaz-Cachinero, J.~I. Muñoz-Hernandez, J.~Contreras, and G.~Muñoz-Delgado,
  ``An enhanced delivery route operational planning model for electric
  vehicles,'' \emph{IEEE Access}, vol.~8, pp. 141\,762--141\,776, 2020.

\bibitem{Zhou_Yang_2019}
Y.~Zhou, A.~Ravey, and M.-C. Péra, ``A survey on driving prediction techniques
  for predictive energy management of plug-in hybrid electric vehicles,''
  \emph{Journal of Power Sources}, vol. 412, pp. 480--495, 2019.

\bibitem{Eider_Markus_2018}
M.~Eider and A.~Berl, ``Dynamic {EV} battery health recommendations,'' in
  \emph{Proceedings of the Ninth International Conference on Future Energy
  Systems}, ser. e-Energy '18.\hskip 1em plus 0.5em minus 0.4em\relax New York,
  NY, USA: Association for Computing Machinery, 2018, p. 586–592.

\bibitem{Lee_Jin_2022}
J.-O. Lee and Y.-S. Kim, ``Novel battery degradation cost formulation for
  optimal scheduling of battery energy storage systems,'' \emph{International
  Journal of Electrical Power \& Energy Systems}, vol. 137, p. 107795, 2022.

\bibitem{Calearo_Degrag_2019}
L.~Calearo, A.~Thingvad, and M.~Marinelli, ``Modeling of battery electric
  vehicles for degradation studies,'' in \emph{2019 54th International
  Universities Power Engineering Conference (UPEC)}, 2019, pp. 1--6.

\bibitem{Schneider_Simon_2021}
S.~F. Schneider, P.~Novák, and T.~Kober, ``Rechargeable batteries for
  simultaneous demand peak shaving and price arbitrage business,'' \emph{IEEE
  Transactions on Sustainable Energy}, vol.~12, no.~1, pp. 148--157, 2021.

\bibitem{Toth_2014_VRP}
P.~Toth and D.~Vigo, \emph{Vehicle routing: Problems, methods, and
  applications}.\hskip 1em plus 0.5em minus 0.4em\relax SIAM, 2014.

\bibitem{Pyomo2017Hart}
W.~E. Hart, C.~D. Laird, J.-P. Watson, D.~L. Woodruff, G.~A. Hackebeil, B.~L.
  Nicholson, and J.~D. Siirola, \emph{Pyomo--optimization modeling in python},
  2nd~ed.\hskip 1em plus 0.5em minus 0.4em\relax {Springer Science \& Business
  Media}, 2017, vol.~67.

\bibitem{MindtPy}
D.~E. Bernal, Q.~Chen, F.~Gong, and I.~E. Grossmann, ``Mixed-integer nonlinear
  decomposition toolbox for {Pyomo} {(MindtPy)},'' in \emph{13th International
  Symposium on Process Systems Engineering (PSE 2018)}, ser. Computer Aided
  Chemical Engineering, M.~R. Eden, M.~G. Ierapetritou, and G.~P. Towler,
  Eds.\hskip 1em plus 0.5em minus 0.4em\relax Elsevier, 2018, vol.~44, pp.
  895--900.

\bibitem{IPOPT}
A.~W\"{a}chter and L.~T. Biegler, ``On the implementation of a primal-dual
  interior point filter line search algorithm for large-scale nonlinear
  programming,'' \emph{Mathematical Programming}, vol. 106, no.~1, pp. 25--57,
  March 2006.

\bibitem{Gebbran_APEN}
D.~Gebbran, S.~Mhanna, Y.~Ma, A.~C. Chapman, and G.~Verbi\v{c}, ``Fair
  coordination of distributed energy resources with {Volt-Var} control and {PV}
  curtailment,'' \emph{Applied Energy}, vol. 286, p. 116546, Mar. 2021.

\bibitem{Jafari_Mehdi_2018}
M.~Jafari, A.~Gauchia, S.~Zhao, K.~Zhang, and L.~Gauchia, ``Electric vehicle
  battery cycle aging evaluation in real-world daily driving and
  vehicle-to-grid services,'' \emph{IEEE Transactions on Transportation
  Electrification}, vol.~4, no.~1, pp. 122--134, 2018.

\end{thebibliography}

\end{document}